# Challenges in Selecting Software to be Reused[*]

Daniel S. Katz, National Science Foundation, dkatz@nsf.gov

(a position paper for:
Sharing, Re-Use and Circulation of Resources in Cooperative Scientific Work – a CSCW'14 workshop)

Software is an integral enabler of computation, experiment and theory and a primary modality for realizing NSF's Cyberinfrastructure Framework for 21st Century Science and Engineering (CIF21) vision[1]. Scientific discovery and innovation are advancing along fundamentally new pathways opened by development of increasingly sophisticated software. Software is also directly responsible for increased scientific productivity and significant enhancement of researchers' capabilities. In order to nurture, accelerate and sustain this critical mode of scientific progress, NSF has established the *Software Infrastructure for Sustained Innovation (SI²) program*, with the overarching goal of transforming innovations in research and education into sustained software resources that are an integral part of the cyberinfrastructure.

SI²'s intent is to foster a pervasive cyberinfrastructure to help researchers address problems of unprecedented scale, complexity, resolution, and accuracy by integrating computation, data, networking, observations and experiments in novel ways. NSF expects that its SI² investment will result in robust, reliable, usable and sustainable software infrastructure that is critical to achieving the CIF21 vision and will transform science and engineering while contributing to the education of next generation researchers and creators of future cyberinfrastructure.

It is expected that SI² will generate and nurture the interdisciplinary processes required to support the entire software lifecycle, and will successfully integrate software development and support with innovation and research. Furthermore, it will result in the development of sustainable software communities that transcend scientific and geographical boundaries. The goal of the SI² program is to create a software ecosystem that includes all levels of the software stack and scales from individual or small groups of software innovators to large hubs of software excellence.

The SI2 program, similar to many other NSF programs, primarily supports projects that are proposed in response to solicitations, such as the recent NSF 13-525[2]. These proposals are then reviewed by a peer-review group, who in addition to the standard NSF criteria of intellectual merit and broader impacts, review the project on the following criteria:

- Does the proposal discuss how the proposed software will fill a recognized need and advance research capability within a significant area (or areas) of science and engineering?
- Does the proposal provide a project plan and timeline, including a proof-of-concept demonstration of any key software element and the steps necessary presented to take the software from prototype to dissemination into the community as reusable software resources?
- Are tangible metrics described to measure the success of any software that may be developed?
- Does the software engineering and development plan include and/or enable the integration of relevant research activities to ensure the software is responsive to new computing developments?
- To what extent are issues of sustainability, manageability, usability, composability, and interoperability addressed and integrated into the proposed software?
- Does the project plan include user interaction, a community-driven approach, and a timeline of new feature releases? Does it plan to extend the work to additional user communities?

Once the peer-review committee has performed its review of the project, NSF must decide which projects should be recommended for funding, based on these reviews and its own judgment how best to balance potential projects vs. total impact across all fields of science and engineering that NSF supports.

---

In both stages (peer reviews and internal NSF deliberations), a key decision factor is predicting scientific impact. Arguments about impact are often made in one or two steps: 1. For both existing and new software, is there a clear need for the software? If it competes with other software, why is it expected that this software will be used, rather than the competing software? 2. If the software already exists, how has it been used in the past?

The first question is often answered either anecdotally or though surveys of potential users. However, the second question can, in theory, be answered more precisely. The software developer wants to answer questions such as: How many downloads are there? How many contributors are there? How many uses have occurred? How many papers cite it? How many papers that cite it are cited? These questions range from easiest to hardest to measure, and from least to most value.

It seems clear that by measuring software usage we add an incentive for good software to be developed and made available, which potentially increases reuse and diminishes redundant development. This need to measure software usage, and to measure researchers' output, is described by Priem et al. in the Altmetrics manifesto[3], by Shiermeier[4], and by many others. When software is used for science, it is also important to credit the software's developers, and since analysis of discoveries is often done through citation[5], I believe we need to create a culture of software citation. A more detailed proposal of how we might do this through transitive credit is available.[6]

My interest in this is two-fold. As a program officer, I want to make the best possible funding recommendations, and as mentioned previously, I want to know about the usage of the software packages that are proposed for potential funding as an input for those recommendations. As a researcher in software intensive areas of computational and data-enabled science and engineering, I want to be sure that I am credited for my software contributions, and that I and my students have career options that recognize and reward our software.

I[7] am a Senior Fellow in the Computation Institute (CI) at the University of Chicago and Argonne National Laboratory and am currently a Program Director in the Division of Advanced Cyberinfrastructure at the National Science Foundation. My interest is in the development and use of advanced cyberinfrastructure to solve challenging problems at multiple scales, including technical research interests in applications, algorithms, fault tolerance, and programming in parallel and distributed computing, including HPC, Grid, Cloud, etc., and policy research interests in citation and credit mechanisms and practices associated with software and data, organization and community practices for collaboration, and career paths for computing researchers.